\documentclass[5p]{elsarticle}
\biboptions{sort&compress}
\usepackage{hyperref}
\hypersetup{
colorlinks=true,
linkcolor=blue,
anchorcolor=blue,
citecolor=blue,
filecolor=blue,
urlcolor=blue
}

\usepackage[utf8]{inputenc}
\usepackage{xcolor}
\usepackage{soul}
\usepackage{comment}
\usepackage{amsmath}
\usepackage{amssymb}
\usepackage{float}
\usepackage{multirow}
\usepackage[colorinlistoftodos]{todonotes}
\usepackage{lipsum}
\usepackage[nameinlink,capitalize]{cleveref}
\usepackage{placeins}
\usepackage{orcidlink}

\usepackage{siunitx}
\usepackage[nolist]{acronym}
\usepackage{caption}
\usepackage{amssymb,amsmath,amsthm,enumitem}
\usepackage{gensymb}
\usepackage{subfigure}
\usepackage{wrapfig}
\usepackage{lscape}
\usepackage{rotating}
\usepackage[normalem]{ulem}
\useunder{\uline}{\ul}{}
\journal{Journal}
\DeclareUnicodeCharacter{FFFD}{fi}

\begin{document}

\begin{frontmatter}

\title{{\bf \LARGE Materials Design for Hypersonics}}


\author[a]{Adam B. Peters*,\orcidlink{0000-0002-3217-7965}}
\author[a]{Dajie~Zhang\,\orcidlink{0000-0003-4763-3692}}
\author[b]{Samuel~Chen\,\orcidlink{0000-0001-7740-006X}}
\author[c]{Catherine~Ott}
\author[a]{Corey~Oses\,\orcidlink{0000-0002-3790-1377}}
\author[d,e]{Stefano~Curtarolo\,\orcidlink{0000-0003-0570-8238}}
\author[c]{Ian~McCue}
\author[g]{Tresa~Pollock}
\author[f,h,i]{Suhas~Eswarappa~Prameela*,\orcidlink{0000-0003-3453-0184}}


\address[a]{Department of Materials Science and Engineering, Johns Hopkins University, Baltimore, MD 21218, USA}
\address[b]{Johns Hopkins Applied Physics Laboratory, Laurel, MD 20723}
\address[c]{Department of Materials Science and Engineering, Northwestern University, IL, USA. 60208}
\address[d]{Department of Mechanical Engineering and Materials Science, Duke University, NC, USA. 27708}
\address[e]{Center for Autonomous Materials Design, Duke University, NC, USA. 27708}
\address[f]{Hopkins Extreme Materials Institute, Johns Hopkins University, Baltimore, MD, USA, 21218}
\address[g]{Departments of Materials Engineering, University of California, Santa Barbara, CA, 93106}
\address[h]{Department of Materials Science and Engineering, MIT, Cambridge, MA, USA, 02139}
\address[i]{Department of Aeronautics and Astronautics, MIT, Cambridge, MA, USA, 02139}


\begin{abstract} 
Hypersonic vehicles must withstand extreme conditions during flights that exceed five times the speed of sound. These systems have the potential to facilitate rapid access to space, bolster defense capabilities, and create a new paradigm for transcontinental earth-to-earth travel. However, extreme aerothermal environments create significant challenges for vehicle materials and structures. This work addresses the critical need to develop resilient refractory alloys, composites, and ceramics. We will highlight key design principles for critical vehicle areas such as primary structures, thermal protection, and propulsion systems; the role of theory and computation; and strategies for advancing laboratory-scale materials to flight-ready components.

\end{abstract}{}

\begin{keyword}
Hypersonics $|$ Materials Design $|$ Extreme Environments $|$ Thermal Protection Systems $|$ High-Entropy Alloys $|$ Ultra High-Temperature Ceramics \\
\vspace{3 mm}

*Corresponding authors: Adam B. Peters (apeter57@alumni.jh.edu) \\
\hspace{39mm}Suhas Eswarappa Prameela (suhasep@mit.edu) 
\end{keyword}

\end{frontmatter}

\onecolumn

\begin{figure*}[ht!]
    \centering
    \includegraphics[width=0.99\linewidth]{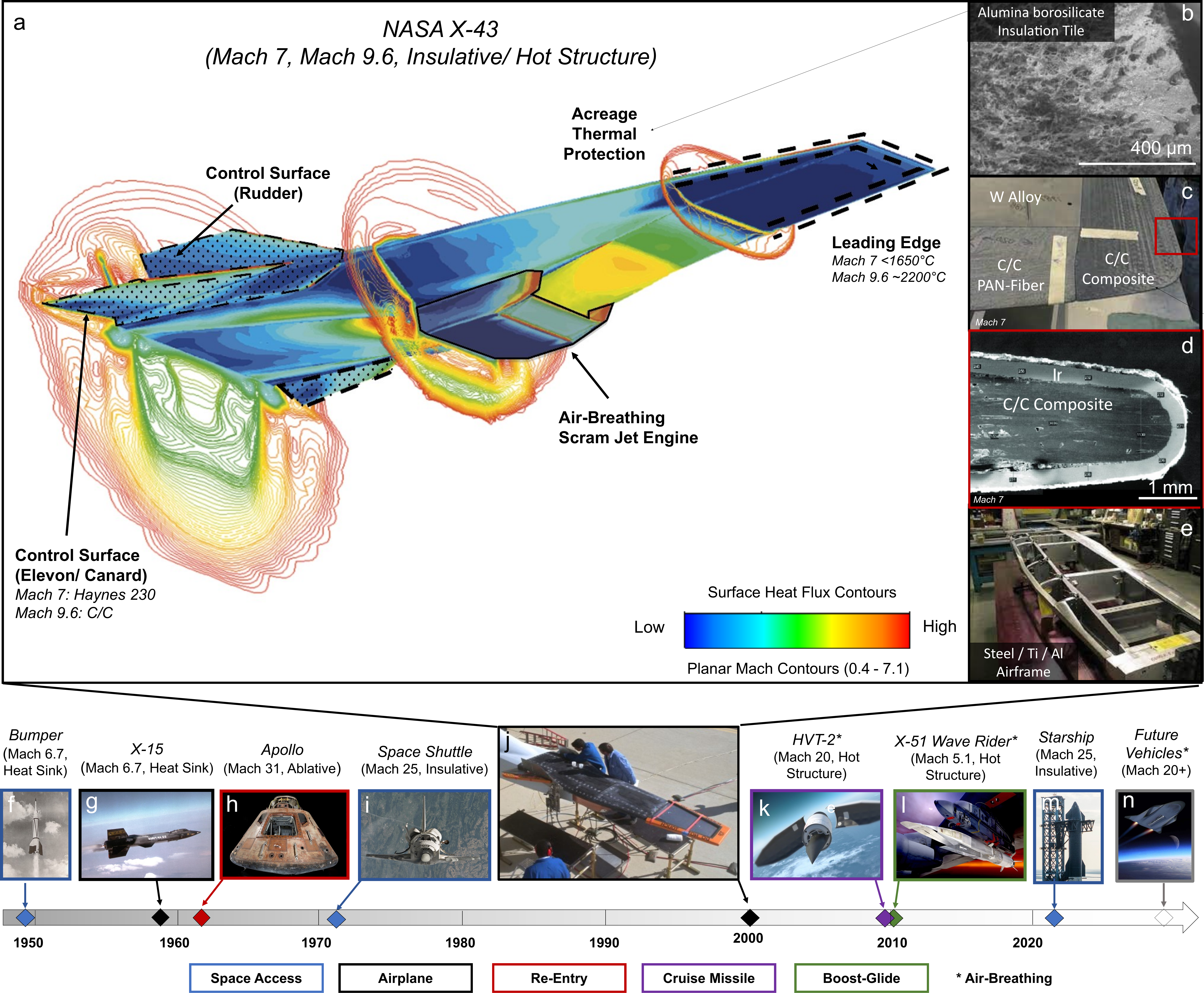}
    \caption{\small {\bf (a)} Computational fluid dynamics (CFD) simulation of the X-43 vehicle at a Mach 7 test condition with the engine operating. The solution includes internal (air-breathing scramjet engine) and external flow fields, including the interaction between the engine exhaust and vehicle aerodynamics. The image illustrates surface heat transfer on the vehicle (red is the highest heating) and flow field contours at the local Mach number. Structural components and the associated materials used for the design of the X-43 hypersonic vehicle are indicated: {\bf (b)} alumina borosilicate insulation tile with an emissive coating used for acreage protection thermal protection~\cite{johnson2012thermal}; {\bf (c)} nose and leading-edge design integrating carbon composites and refractory tungsten alloy SD 18; {\bf (d)} sharp leading edge cross-section showing the carbon composite with a refractory Ir coating~\cite{glass2008ceramic}; {\bf (e)} airframe of the vehicle composed of steel/aluminum skin and Al/Ti bulkheads. {\bf (f-n)} Timeline of hypersonic vehicle development spanning hypersonic airplanes, space access, re-entry, boost-glide vehicles, and cruise missile applications, where colors indicate the hypersonic vehicles configuration: {\bf (f)} the first to reach hypersonic speeds, Project Bumper-WAC ``Without Any Control'' (1949), {\bf (g)} the reusable X-15 research aircraft (1959), {\bf (h)} Apollo re-entry capsules (1961-1972), {\bf (i)} Space Shuttle (1972-2011), {\bf (j)} NASA X-43 airplane (2001), {\bf (k)} the HVT-2 boost-glide vehicle (2010-2012), {\bf (i)} Boeing X-51 scramjet (2010-2013), {\bf (m)} SpaceX Starship (slated for hypersonic re-entry in 2023), {\bf (n)} a notional future hypersonic airplane. (Image sources: NASA (a,b,c,d,e,f,h, i,j), (c) – adapted from~\cite{ohlhorst1997thermal}, (d) – adapted from~\cite{glass2008ceramic}, U.S. Airforce (g,l), DARPA (k), Creative Commons - Offical SpaceX (m), U.S. Govt. images not subject to copyright)}
    \label{fig:1}
\end{figure*}

In the last decade, there has been a resurgence in hypersonic vehicle development driven by the desire to increase flight performance and reusability. Hypersonics refers to flight and aerodynamic phenomena that occur above Mach 5 (5 times the speed of sound). To frame hypersonic speeds, a non-stop flight from Los Angeles to Tokyo aboard a commercial airliner (Mach 0.8) takes roughly twelve hours, whereas onboard an emerging Mach 9 hypersonic vehicle it takes one.  Although the first hypersonic flight was achieved ${\sim}$70 years ago, there has been increasing interest from a broader audience due to modern engineering advances that are poised to revolutionize defensive capabilities, sub-orbital travel, and rapid access to space~\cite{van2021hypersonics, voland2006x, deminsky2010hypersonics} (\textcolor{blue}{Figure} \ref{fig:1}). Candidate vehicle systems with ever-increasing capabilities and Mach numbers are being developed, including boost-glide systems, reusable aircraft, space-launch vehicles, and missile technologies~\cite{van2021hypersonics}. However, these remarkable leaps in Mach number and performance during atmospheric flight come with an array of formidable challenges in the domain of materials multi-property optimization, simulation, and design~\cite{EswarappaPrameela2023}. Vehicles are purpose-built with bespoke materials to operate at vastly different Mach numbers (5-25+), altitudes (spanning sea level to orbit), hypersonic flight times (ranging from seconds to hours), and trajectories. \\ 

When vehicle speeds increase past supersonic conditions and into the hypersonic regime, the physics of external aerodynamic flows become dominated by aerothermal heating rather than aerodynamic forces (\textcolor{blue}{Figure} \ref{fig:1}\textcolor{blue}{a}). Aerodynamic compression and friction create high-enthalpy gas dynamics that impart additional physical phenomena from the energy exchange of a superheated atmosphere. This superheated atmosphere results in: high heat fluxes (3-7 orders of magnitude greater than the 1.4 kW/m$^2$ from the sun); extreme thermal gradients (changing from -170$^\circ$C to 3000$^\circ$C across distances of order 1 cm); high stagnation pressures (${\sim}$ 10$^5$-10$^7$ Pascals); and destructive plasma from gas ionization which accelerates materials oxidation~\cite{van2021hypersonics, glass2008ceramic, chen2020modeling}. As operational Mach numbers increase, these formidable phenomena must be accommodated by materials in the principal subsystems of a hypersonic vehicle: aeroshell/primary structure, leading edges, control surfaces, acreage thermal protection, propulsion, and guidance systems. Existing hypersonic materials limit the resiliency of structures during operation in extreme environments. Improving such materials has become the focus of cutting-edge research.  \\ 

Materials for hypersonics can be broadly classified into three types: refractory metals, composites, and ceramics. Recent work has focused on their development for propulsion systems, ~\cite{huda2013materials,meng2021micromanufacturing,qu2018microstructural} thermoelectric generators,~\cite{gong2018novel}, radomes~\cite{khatavkar2016composite}, structural materials~\cite{marshall2014national}, and thermal protection systems~\cite{marshall2014national, li2021improved}. Each material system offers distinct tradeoffs for a given sub-system and environmental application. Common metals and alloys in hypersonics, such as aluminum and nickel-base superalloys, are favorable for primary structural components and moderate thermal loads ($<$800\degree C), while refractory metals with higher operating temperatures (800-1200\degree C) are employed for structures that see more demanding operating conditions in oxidizing atmospheres(\textcolor{blue}{Box 1}).  Refractory ceramics combine high-temperature capability (\textgreater 1700°C) with moderate thermal conductivity, but lack monolithic thermal shock resistance and tend to be used as a thermal barrier coating or thin structural materials~\cite{glass2011physical}. By contrast, fiber-reinforced composite materials, such as carbon/carbon or ultra-high temperature ceramics matrix composites, incorporate carbon or ceramic fibers in dense matrices to improve high-temperature strength-to-weight ratios beyond metals~\cite{glass2008ceramic,glass2011physical, glass2018thermal, glass2006materials}. \\ 

Advancement of these materials from laboratory scale studies to flight is hindered by standardization of materials processing; reproducibility of materials data; and difficulties testing representative thermal, oxidative, and mechanical flight conditions. High-fidelity models have historically been used to design flight trajectories to bound materials selection criteria. Advanced materials design tools are emerging with integrated computation and predictive frameworks that can aid the design of complex materials and expand vehicle performance and reliability. We will explore how refractory metals, composites, and ceramics are designed and selected for hypersonic applications according to vehicle-specific design criteria. 

\section*{\large{Hypersonic Vehicle Configurations and Design Requirements}}
Material requirements for hypersonic flight are sensitively coupled to the vehicle design and flight envelope, which impose two principle environmental challenges: (1) thermal loads that are dependent on both geometry and location on the vehicle; (2) strongly oxidizing conditions that drive changes in both material properties (oxidation) and geometry (ablation). As a result, aerostructures, wing leading edges, acreage thermal protection systems, and propulsion systems necessitate vastly different materials to accommodate these diverse thermo-chemo-mechanical loads. Depending on the flight conditions (Mach and altitude), the flight time at a given Mach number and altitude (known as time on condition), and location on the vehicle, qualified materials may not exist for the desired application~\cite{marshall2014national}. \\ 

Aerothermal heating arises as the hypersonic vehicle pierces through the atmosphere. Fundamentally, the adiabatic dissipation of a vehicle’s kinetic energy into the viscous gas environment is responsible for the extreme thermal conditions of flight~\cite{smith2021aerodynamic}. In the vehicle's shock layer, (volume gas between the body and the shock wave), stagnation temperature increases proportionally to Mach to the third power and root of the atmospheric density and can reach values as high as 10,000$^\circ$C~\cite{allen1964aerodynamic}. Although much of the energy is swept away with the surrounding gas flow around the vehicle, energy transfer by convective or radiative heating generates high heat fluxes that necessitate materials capable of resisting high temperatures~\cite{huda2013materials}. \\

Material requirements are further exacerbated due to the dissociation of O$_2$ and N$_2$ into free radicals at gas-phase temperatures above 3000$^\circ$C (typically at speeds greater than Mach 8). These conditions lead to highly reactive surface chemical interactions, causing materials degradation, microstructural evolution, phase formation, and property changes during flight~\cite{chen2020modeling, marshall2014national, chaudhry2020vehicle}. Critical challenges for materials designers remain in both the leading-edge surfaces from direct aerothermal exposure (nose, cowl lips, and control surfaces, \textcolor{blue}{Figure} \ref{fig:1}\textcolor{blue}{a}), and in the propulsion flow path where radiative cooling is not viable~\cite{marshall2014national}. In the following section, we will highlight how the characteristics of the primary aerostructures, thermal protection systems, and propulsion systems influence material design and selection.\\

\textbf{\small{Primary Aerostructure}.}
\label{ssec: 2.1}Lightweight primary structures (e.g., aeroshells and airframes) may be formed into either lifting bodies (an aircraft or spacecraft configuration that produces lift) or ballistic structures (elements that relies on projectile motion), where the leading-edge profile and flight trajectory govern the aerothermal load during flight. Unlike the traditional atmospheric re-entry vehicle designs in \textcolor{blue}{Figure} \ref{fig:1}\textcolor{blue}{h} – which employ blunt features to increase drag and push the shock region away from the structure and transfer energy into the air – hypersonic vehicles require slender primary structures and sharp control surfaces to reduce drag and enable stable long-distance accuracy. However, the heating rate is inversely proportional to the square root of the tip radius and must be accommodated through various energy dissipation mechanisms. \\

In modern vehicles, aeroshells are designed using solid or sandwich constructions with honeycomb, lattice, corrugated, or foam cored to minimize weight while maintaining rigidity and enable advanced passive cooling strategies~\cite{glass2008ceramic, glass2018thermal, glass2006materials, ohlhorst1997thermal, krenkel2005carbon, tenney1989materials}. Robust carbon and ceramic composites remain materials of choice for modern leading-edge structures~\cite{glass2008ceramic, glass2018thermal, glass2006materials, ohlhorst1997thermal, krenkel2005carbon, tenney1989materials}, and enable peak temperature reduction through passive cooling by employing favorable composite weave patterns, or thermally conductive materials to more effectively transport heat to the colder regions of the aeroshell main body~\cite{glass2006materials, krenkel2005carbon}. Such designs are commonly referred to as ``hot structures'' (\textcolor{blue}{Figure} \ref{fig:1}\textcolor{blue}{k, l}) as compared to the insulated ``cold structure'' design adopted by the Space Shuttles and many other types of reentry vehicles or bodies that use thick outer surface thermal insulation (\textcolor{blue}{Figure}  \ref{fig:1}\textcolor{blue}{i,m}). \\

\textbf{\small{Thermal Protection System}.} Thermal protection materials and system design has become an engineering field of its own because materials with unique property combinations can enable previously unmatched flight capabilities. Thermal protection systems (TPS) are employed for thermal regulation of leading edges, nose, and propulsion features that experience the greatest heat flux, as well as acreage locations that protect the aeroshell’s fuselage and control surfaces (i.e., rudders, elevons; \textcolor{blue}{Figure} \ref{fig:1}\textcolor{blue}{a}). In modern vehicles, the aerostructure may have an integrated TPS for optimal heat transfer and dissipation. TPS materials are selected to best accommodate the local aerothermodynamic criteria according to their combinations of high-temperature strength, thermal conductivity, heat capacity, melting/oxidation temperature, and emissivity. Broadly, there are three fundamental types of TPS used to increase vehicle resilience to aerothermal heating: passive, semi-passive, and active TPS~\cite{glass2008ceramic, marshall2014national},~\cite{glass2011physical, glass2018thermal, glass2006materials}. \\ 

Passive thermal protection systems are ideal for moderate transient heat flux scenarios and may be composed of (i) insulated cold structures (e.g., Space Shuttle tiles); (ii) heat sink surface structures that both absorb and radiate energy (e.g., the skin of the X-15); or (iii) or emissive ``hot structures'' that lower the thermal load through both environmental radiation and conduction into the vehicle (e.g., the nose of the X-51). Examples of these appear in \textcolor{blue}{Figure} \ref{fig:1}\textcolor{blue}{g, i}. Semi-passive systems are implemented for high heat fluxes that persist for long durations and encompass (i) reusable heat pipes that transfer and radiate thermal energy via evaporative cooling and capillary wicking (e.g. liquid lithium or potassium~\cite{fusaro2022liquid});  or (ii) single-use ablatives materials that absorbed energy via pyrolysis or charring of a reinforced polymer/resin (used in the first re-entry capsules, \textcolor{blue}{Figure} \ref{fig:1}\textcolor{blue}{h}). \\ 

Active thermal protection systems, using the forced flow of a liquid or vapor, are employed for the most extreme heat fluxes and extended flight durations~\cite{gross1961,holden1994,naved2023,liu2010,su2019, ifti2022laminar}. These systems include (i) convective cooling architectures, which transfer heat into a working fluid (e.g., Shuttle main engine), (ii) film cooling, whereby a fluid is injected over a large area into the flow to form an insulating blanket (e.g., X-43 propulsion system), (iii) or transpiration cooling where a fluid (e.g., H$_2$O or He) is injected into hot gas flow through porous structures. Examples of these appear in \textcolor{blue}{Figure} \ref{fig:1}\textcolor{blue}{a, i, j}. \\

Nose and wing leading edges that are subjected to intense heat loading may employ heat pipes or actively cooled structures for thermal regulation~\cite{marshall2014national, frey2022high}. By contrast, acreage locations — large fuselage regions on the vehicle — experience lower heat fluxes and have historically been passively cooled using materials with low thermal conductivities and thermal expansion coefficients (\textcolor{blue}{Box 1}). Although heat sinks and thermal insulation are attractive from a risk management perspective, they suffer from excessive mass and low fracture toughness. Ablative materials have been used to great effect for shuttle re-entry, with recent developments centered on enhancing performance by employing graded structures and unique compositions achieved through additive manufacturing, still, ablative thermal protection systems do not favor re-usability ~\cite{olson}. In contrast, hot structures, heat pipes, and active thermal management systems (enabled by additive manufacturing) have dominated research efforts for candidate materials and system designs~\cite{Naved_2023, Ewenz_Rocher_2022, Reimer_2011, Ravichandran_2023, HOLDEN_1990, Liu_2010, Su_2019, Steins_2022}. \\

\begin{figure*}[ht!]
    \centering
    \includegraphics[width=0.99\linewidth]{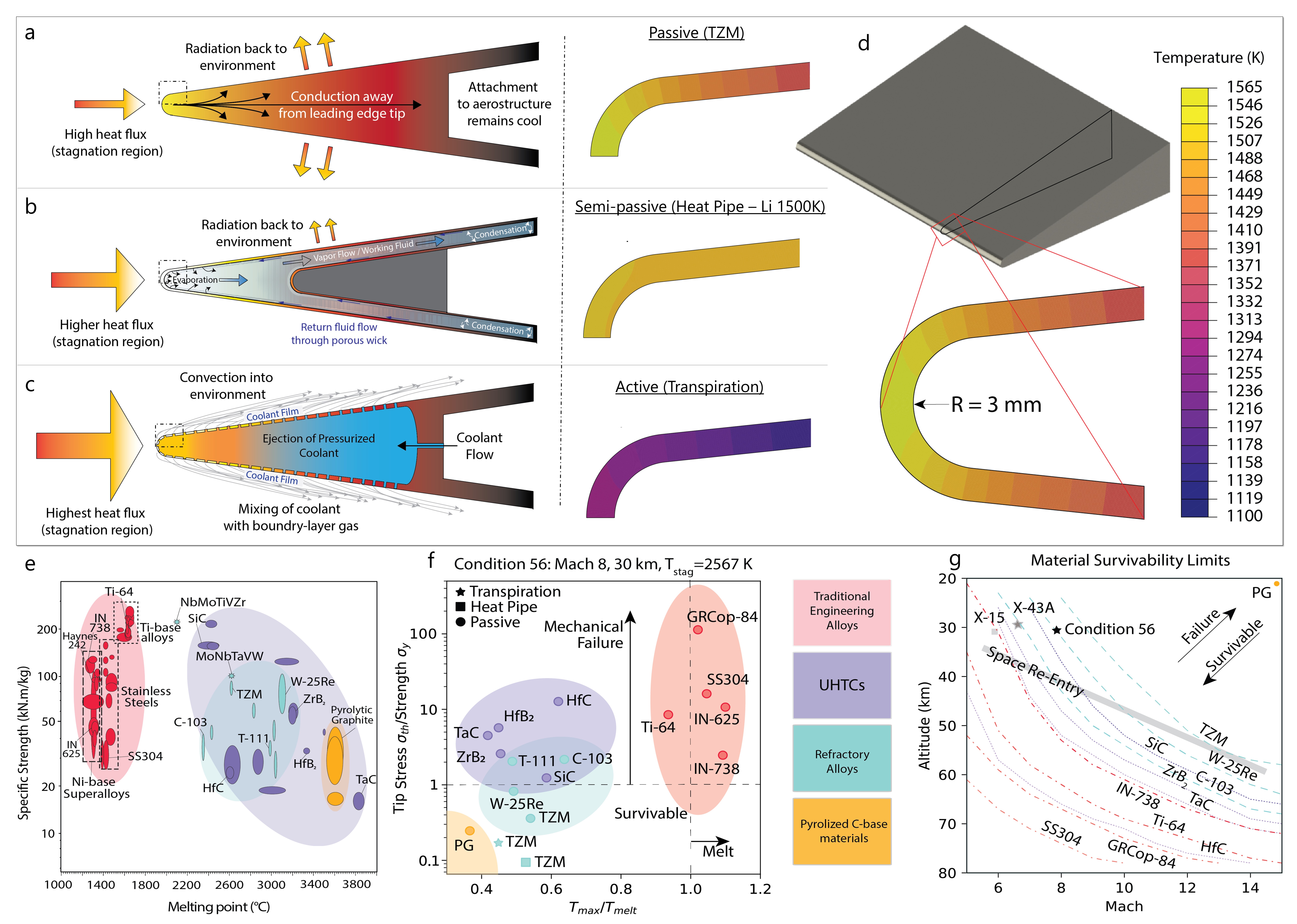}
    \caption{\small \textbf{Leading edge thermal protection systems types and steady-state finite element (FE) simulations of aerodynamic heating of a leading edge, carried out for a range of structural materials and hypothetical hypersonic flight conditions.} {\bf (a)} Illustration of the passive leading edge (left) and thermal profile (right) across a 2D TZM leading-edge, considering passive thermal management; (b) illustration of semi-passive leading edge (left) and thermal profile (right) of a semi-passive Li heat pipe operating at 1500 K; {\bf (c)} illustration of active leading edge (left) and thermal profile (right) showing transpiration where the incident heat flux is reduced by a factor of 2. {\bf (d)} Illustration of the sharp leading-edge geometry used in these simulations with the following dimensions: 3 mm tip radius, 3-degree wedge angle, 5 cm span, 10 cm cord length. {\bf (e)} Ashby map highlighting operational tradeoffs for metal alloy, UHTC, refractory alloy, and carbon-base material classes.{\bf (f)} Ashby-style plot of the FE simulation results from passive leading edges, where: the y-axis is the normalized mechanical stress resulting from a thermal expansion gradient, and the x-axis is the normalized peak temperature at the tip where the heat flux is highest. Only 4 materials are not viable from a temperature standpoint (ignoring oxidation), whereas 8 are not viable due to the expansion stress exceeding the yield strength of the material at that temperature. Constraints via oxidation will decrease the overall maximum operating temperature; there is limited availability of oxidation kinetics for these materials. {\bf (g)} The culmination of {\bf (f)} for different flight conditions is shown as a hypothetical hypersonic flight corridor, where each line represents the ``survivability limit'' for a monolithic material with this specific (sharp) wedge geometry; known flight conditions of the X-43A, X-15, and typical space re-entry are indicated for reference.}
    \label{fig:2}
\end{figure*}

Each TPS will have a unique architecture and thermal profile, which results in its set material property requirements. An example of simulated aerodynamic heating for a passive (hot structure), semi-active (heat pipe), and active (transpiration cooling) cooled leading edge is shown in \textcolor{blue}{Figure} \ref{fig:2}\textcolor{blue}{a, b} at one flight condition. In this example, each leading edge experiences the same heat flux and stagnation temperature because these are dictated by both the component geometry and flight condition. However, the resulting temperature profile depends on the TPS mechanism. The passive leading-edge exhibits the highest peak temperature and thermal gradient because it relies solely on intrinsic material properties (conductivity, heat capacity, and emissivity). The semi-passive leading-edge exhibits a small thermal gradient (but a similar peak temperature to the passive structure) because heat pipes increase thermal conductivity by 1-3 orders of magnitude~\cite{schlitt1995performance}. Lastly, the actively cooled leading edge has the lowest peak temperature because transpiration reduces the incident heat flux~\cite{eckert1958mass, gross1961review, holden1994experimental, dunavant1969exploratory}. \\

\textbf{\small{Air-Breathing Propulsion Systems}.} Similar to environmentally facing thermal protection systems, existing approaches to hypersonic propulsion systems can be significantly improved with refractory materials capable of operating in stressing aerothermal oxidizing and reducing environments~\cite{van2005hypersonic}. Currently, the ram/scramjet engine is the standard form of propulsion for air-breathing hypersonic vehicles. Unlike rocket-propelled hypersonic vehicles (e.g., X-15 and the Space shuttle), the oxidizer for the propellent is supplied by the surrounding air and mixed within a combustor using onboard fuel~\cite{urzay2018supersonic}.  More advanced combined cycle multi-mode propulsion systems under development include rocket-based combined cycle (RBCC); turbine-based combined cycle (TBCC); and turbo rocket combined cycle (TRCC) that are capable of transition propulsion modes (such as rocket propulsion during the initial ascent phase and then transition to air-breathing scramjet engines at hypersonic speeds \textcolor{blue}{Figure} \ref{fig:1}\textcolor{blue}{a}). \\ 

Specific components in these propulsion systems, including inlet ducts, nozzles, and combustors, experience extreme temperature and mechanical stress without the ability to readily dissipate heat through radiative cooling. Propulsion system materials include a combination of refractory alloys, CMCs, C/C, and metal matrix composites (MMC)~\cite{suhrutha2020recent}. Textile-based CMCs may be formed into complex structures for internal coolant flow and mechanical stiffening, but their surface temperatures are limited to ${\sim}$1600$^\circ$C. For velocities near Mach 6, passively cooled refractory materials can be used to operate at temperatures near that of the propulsive flow, but active cooling is required above Mach 6.  Active cooling methods require materials that can accommodate high temperatures, pressure, and temperature gradients between the cooled fuel and combustion chamber~\cite{marshall2014national}. \\

Materials lifetimes in this environment are largely controlled by oxidation, which is highly dependent on flow conditions when water vapor is present in the propulsion flow – this oxidation mechanism is quite different from the ionized flow in leading-edge applications. Oxidation is exacerbated by thermal gradient-induced microcracking, and limitations in high-fidelity modeling make materials properties insufficient for lifetime prediction~\cite{marshall2014national, ohlhorst1997thermal, krenkel2005carbon}. These futuristic propulsion systems could enable low-cost and reusable air-breathing hypersonic vehicles for manned flights and civilian transportation, but materials development is necessary. More efficient materials will require accurate prediction of engine thermal balance, heat loads, shock conditions, and oxidative character of the burning atmosphere~\cite{van2005hypersonic}.


\begin{sidewaysfigure*}
    \centering
    \includegraphics[width=0.99\linewidth]{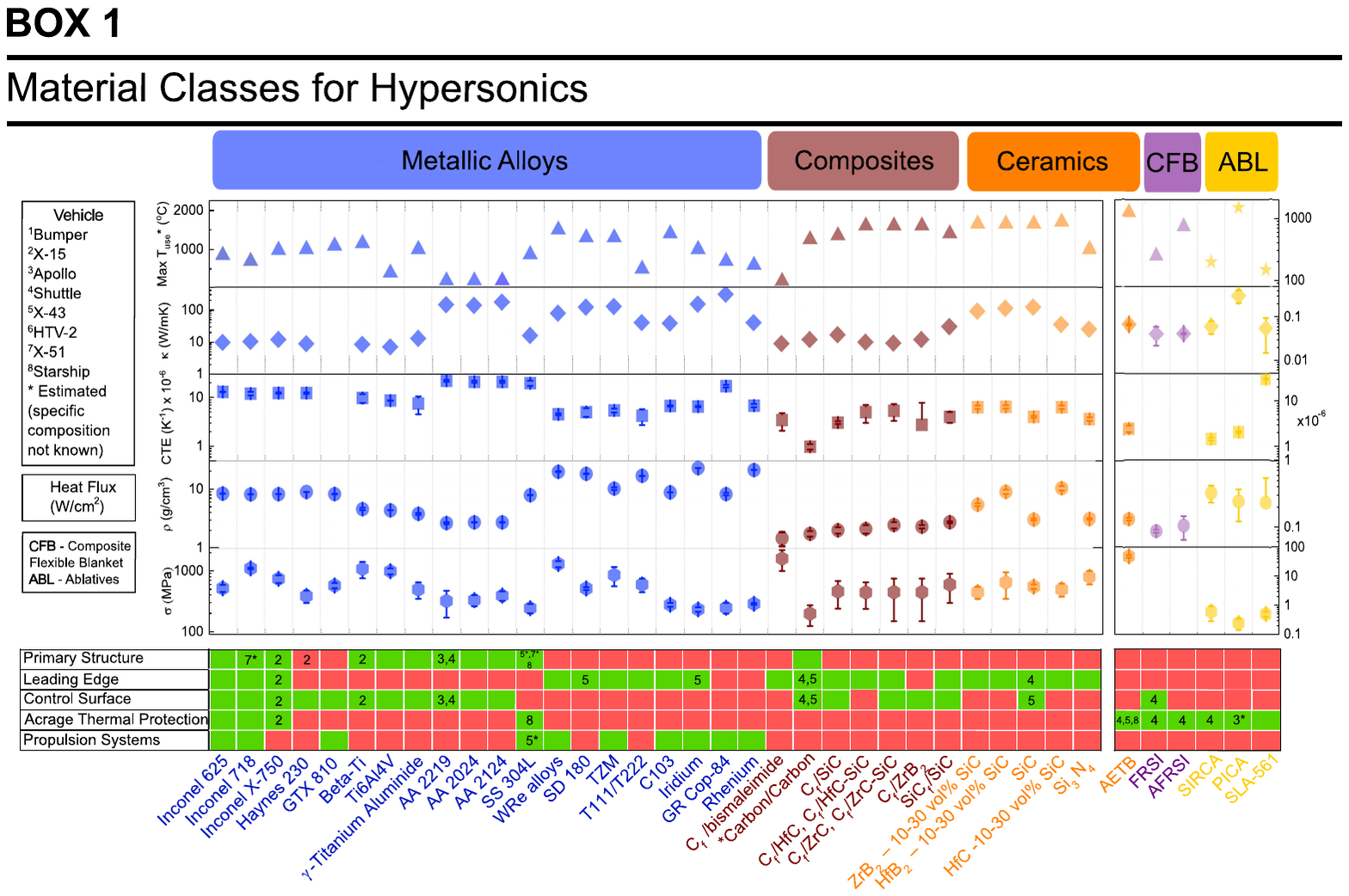}
    \vspace*{-15mm}
    \caption*{\small \textbf{Box 1:} \textbf{Material Classes for Hypersonics}. Materials properties such as tensile strength ($\sigma_{TS}$), yield strength ($\sigma_y$), flexural strength ($\sigma_{FS})$, density ($\rho$), coefficient of thermal expansion (CTE), thermal conductivity ($\kappa$), various use temperatures (T), and heat flux (Q) are given for metallic alloys, composites, ceramics and synthetics. Some materials data are not available due to limited availability. Some of the variation in the material's properties is highly variable based on weave, composite structure, and processing. The numbers in boxes correspond to the materials used on the vehicles depicted in Figure \ref{fig:1}. Common material property data given from~\cite{hartleibtpsx, matwebOnlineMaterials}}
    \label{fig:table}
\end{sidewaysfigure*}


\section*{\large{Materials-class-specific Considerations and Design Criteria}}
Materials selection is typically applied after structural components have been designed and trajectories have been determined~\cite{stort2023}. Initial material screening can be carried out using thermo-mechanical simulations. For a given set of material properties, conditions (heat flux and stagnation temperature) are applied across a component to calculate the resulting thermal profile, which is then used as boundary conditions to calculate thermal stresses. This screening is useful in determining whether the peak temperature exceeds a material’s melting point and/or the thermal stress exceeds the material’s flow stress at the given temperature. \\

An example of this screening is illustrated in \textcolor{blue}{Figure} \ref{fig:2}\textcolor{blue}{c-e} for a sharp passive leading edge. As a hypothetical evaluation, steady-state thermal simulations were carried out on high-temperature structural materials for over 300 Mach and altitude combinations (i.e., assuming the material was exposed to these conditions indefinitely and allowed to equilibrate). For each simulation, following the work of~\cite{steeves2009influence}, we extracted the peak temperature and estimated the tip stress from the thermal gradient in this region. The incident heat flux changes substantially around the tip, \textcolor{blue}{Figure} \ref{fig:2}\textcolor{blue}{b}, causing a steep thermal gradient (as large as 1000 K across 4 mm), which generates stresses of order 100 MPa due to non-uniform thermal expansion. If these stresses are above the materials’ flow stress at the peak temperature, the tip will deform and affect the boundary layer, potentially causing a laminar-to-turbulent transition\\

These mesoscale models provide critical insights into guiding what materials should be used in the various structures (mentioned in the previous section) for a given set of flight conditions. For instance, \textcolor{blue}{Figure} \ref{fig:2} highlights that traditional alloys (e.g., Ti-base, Ni-base and steels) have limited use as leading edges due to low melting points, high-thermal expansion coefficients and moderate thermal conductivities (\textcolor{blue}{Figure} \ref{fig:2}\textcolor{blue}{d}). Monolithic ceramics suffer from high thermal stresses, but their strengths can be modified through secondary phases (see Ultra-high Temperature and Refractory Ceramics for Hypersonics) or employed as coatings. Refractory metals – owing to their high strength at temperature, thermal transport properties, and low thermal expansion coefficients – are ideal, but oxidation kinetics will constrain their maximum service temperature (see Metallic Materials for Hypersonics).  \\

\textbf{\small{Metallic Materials for Hypersonics}.} Metallic materials are ubiquitously used in hypersonic vehicles – as nose and wing leading edges, control surfaces, and engine inlets – due to their tendency for damage tolerance and manufacturability. These components need to withstand extremely high heat fluxes and thermal strains, which demand materials with high melting points that maintain strength at high temperatures. \\

Pure elements with high melting points (W, Re, Ta, Mo, Nb, V, Cr, Ti, Ni) form the basis of fielded high-temperature alloys. For instance, Ti was employed in hot aeroshell structures in the SR-71~\cite{glass2008ceramic, glass2018thermal} the nose section of the X-43 contained a SD 180 tungsten heavy alloy~\cite{nasaAdvancesHotStructure}, a Haynes Ni-base alloy was used in the Mach 7 X-43 variant~\cite{usaf2015}, and both MoRe and Ni-base alloys have been tested for heat pipe structures~\cite{khatavkar2016composite}. Other refractory metals, such as the TaWHf alloys T111 and T222 (\textcolor{blue}{Box 1}) exhibit favorable creep-resistant materials properties and are ideal for the extended containment of heated liquid alkali-metal working fluids (1000-1300$^\circ$C) for heat pipe type leading edge designs~\cite{afb1968return}. T111/Li and niobium-based C-103/Na designs have been assessed to satisfy the requirements for Mach 8 and Mach 10 flights respectively~\cite{fusaro2022liquid}.   \\

However, melting point alone is not a singularly meaningful parameter of structural design. For comparison, carbon-carbon will not melt at 1 atm but will sublime at 3727$^\circ$C and oxidize to CO$_{(g)}$ at temperatures starting as low as ${\sim}$370$^\circ$C. As in all flight applications, density, oxidation resistance and the ability to tolerate thermomechanical loading are important considerations (\textcolor{blue}{Figure} \ref{fig:2}\textcolor{blue}{c}). Nickel alloys are capable of operating at high fractions of their melting points due to both coherent precipitation strengthening and self-passivation that persist to very high temperatures.  Meanwhile, W and Mo can maintain over 50\% of their Young’s Modulus at 2000$^\circ$C but oxidize well below 1000$^\circ$C, with more rapid oxidation at increasing temperature due to the high vapor pressures of their trioxides~\cite{glass2008ceramic, li2021improved, glass2011physical}.  The combined property advantages in Ni- and Co-base alloys are often not found in state-of-the-art refractory alloys, inhibiting their true operational potential. \\

One promising class of metallic materials is the newly developed multi-principal element (``high-entropy'') alloys~\cite{cantor2004microstructural, yeh2004nanostructured, calzolari2022plasmonic, senkov2010refractory, borg2020expanded, george2019high}. In particular, refractory multi-principal element (RMPE) alloys offer the opportunity to maintain high-temperature properties while, for example, decreasing density and oxidation kinetics. RMPE alloys, such as MoNbTaVW ($\sigma_y$=1246 MPa, $\rho$=12.4 g/cm$^3$)~\cite{senkov2010refractory} and NbMoTiVZr ($\sigma_y$=1785 MPa, $\rho$=7.1 g/cm$^3$)~\cite{zhang2012alloy}, could provide significant benefits over legacy refractory alloys such as Ta-10W, which were chosen for re-crystallization behavior rather than strength ($\sigma_y$=471 MPa, $\rho$=16.8 g/cm$^3$)~\cite{lassila2002effect}. Unfortunately, many thermophysical properties and operation-relevant properties, such as recrystallization temperature, have yet to be measured for many of these alloys. While multidisciplinary design approaches have successfully been implemented for the aerothermal and mechanical design of hypersonic vehicles~\cite{bowcutt2001multidisciplinary}, materials have yet to be factored into this dynamic design optimization loop. \\

Due to their limited oxidation resistance, alloys in hypersonic environments typically rely on a compatible coating. Coatings may be multilayered, functioning as both thermal and environmental barriers, with oxide-forming metallic layers and porous, low-conductivity ceramic overcoats~\cite{pollock2012multifunctional}. Nickel alloys are designed to form alumina as a protective oxide and have well-developed aluminide coating systems due to their extensive use in aircraft engines~\cite{pollock2012multifunctional}.  However, coatings are much less developed for refractory alloys and typically contain metal silicides, which have limited protection below 850 $^\circ$C and slough off above 1700$^\circ$C due to aeroshearing~\cite{perepezko2015surface, SWADZBA2019331}. A wide range of potential coating failure modes need to be considered during design, and while there has been considerable progress on understanding the mechanics of coatings~\cite{borg2020expanded, begley_hutchinson_2017}, often, the material properties are missing. \\

Out of the numerous legacy refractory alloys developed by NASA, only a handful are manufactured and used today (such as C103, TZM, and W25Re). In the past, cost and formability at ambient temperatures were the roadblocks in fielding superior refractory alloys. For instance, Nb alloys with high fractions of W and Hf (greater than 15 wt.\%) suffer from ductile-to-brittle transitions several hundred degrees above room temperature~\cite{SENKOV2019151685}. However, rapidly evolving additive manufacturing capabilities can now produce complex component designs – negating machining constraints~\cite{panwisawas2020metal, livescu2018additively, talignani2022review, higashi2020selective}. This provides opportunities for the design of advanced cooling methodologies such as active TPS~\cite{panwisawas2020metal, livescu2018additively, talignani2022review, higashi2020selective}.  These emerging manufacturing techniques can also be combined with high-throughput characterization and machine learning algorithms to rapidly discover and develop the next-general refractory alloys for hypersonics~\cite{frey2022high,zhang2011mechanical,liu2022machine}. \\

\textbf{\small{Carbon Composites for Hypersonics}.} Carbon-carbon composites (C/C) are historically considered the de facto materials for the fabrication of hypersonic aeroshells and leading edges owing to their excellent performance characteristics, including: low density (1.60-1.98 g/cm$^3$), low coefficient of thermal expansion (-0.85 to 1.1 x 10$^-6$/K), high modulus of elasticity (200 GPa), high thermal conductivities (${\sim}$4-35 W/mK) and retained mechanical properties up to ${\sim}$2000$^\circ$C in inert environments~\cite{glass2008ceramic, marshall2014national,glass2006materials, ohlhorst1997thermal, krenkel2005carbon},~\cite{luo2004thermophysical, ohlhorst1997thermal, buckley1993carbon}. Low-density C/C is often preferred over metals for severe environment aerostructure elements. For instance, the horizontal Haynes control surfaces of the Mach 7 variant of X-43 were replaced with coated C/C for the Mach 10 variant~\cite{ohlhorst1997thermal}. \\

C/C is manufactured using one of two processes to densify a 2-D ``zero condition'' C/C preform: (1) infiltration and pyrolysis (PIP) in which a high carbon yield resin (phenolic or pitch) is infiltrated into the fibrous pre-form (${\sim}$40-60 vol\% fiber) before undergoing high-temperature graphitization of the resin matrix; or (2) chemical vapor infiltration (CVI), where densification occurs by the infiltration and decomposition of carbonaceous gases. Both manufacturing methods have repeated steps to consolidate the matrix, increase density, and increase strength. For PIP, 4-6 pyrolysis cycles under hot isostatic pressure (HIP) are preferred to inhibit the formation of closed pores during resin pyrolysis and achieve a high density ($>$ 98\%). On the other hand, surface carbon deposition via CVI inhibits further infiltration over time so the surface deposit needs to be removed and the process restarted. Advanced C/C 6 (ACC-6) remains the state-of-the-art composite, having undergone six impregnation cycles to increase density and achieve high yield strengths at elevated temperatures. \\

Porosity reduction is a focal point of processing research because it is critical in limiting oxygen diffusion and aerothermal erosion during flight. Uncoated C/C formed by PIP and CVI have demonstrated in-plane tensile strengths on the order of 165 MPa (with strength increasing as a function of density)~\cite{hatta2004tensile}. Furthermore, the thermomechanical properties of C/C are highly anisotropic and dependent on the processing method, residual porosity, and fiber architecture (fiber-woven fabrics or fiber tow windings may be orientated at an angle, such as 30$^\circ$, 60$^\circ$, or 90$^\circ$).  Given this complexity, the need for high-fidelity modeling capabilities for anisotropic, and locally varied, volume elements is critical, which makes materials properties standards insufficient for design, performance, and life prediction~\cite{marshall2014national}. \\

Despite these promising properties, uncoated C/C erodes rapidly at elevated temperatures. The oxidation of carbonaceous composites begins ${\sim}$370$^\circ$C in air, with dramatic oxidation occurring beyond 500$^\circ$C~\cite{li2021improved}. Present hypersonic materials design efforts aim to protect C/C from high-temperature oxidation, ablation, and erosion from prolonged and repeated aerothermal exposure.  With increasingly more extreme hypersonic environments, two protection approaches are being developed: (1) deposition of high-temperature protective coatings, and (2) modification of the carbon-carbon matrix. Oxidation-resistant coating materials may be applied to C/C surface (or fibers before infiltration) to limit diffusion and modify emissivity for passive thermal regulation. \\

The initial development of anti-oxidation coatings for C/C was focused on refractory compositions containing SiC, HfB$_2$, and ZrB$_2$ (HfB$_2$/SiC and ZrB$_2$/SiC blends)~\cite{fu2022micro}. Other additives such as tetraethyl orthosilicate have been applied as silica-forming impregnates (i.e. on Shuttle Orbiter), which seal microstructural defects to limit oxidization~\cite{curry2000orbiter}. These systems can protect C/C from oxidation up to 1500-1600$^\circ$C, but thin oxide coatings become ineffective at higher temperatures due to melting, evaporation/active oxidation of SiO$_{(g)}$, foaming, and/or visco-elastic erosion of the HfO$_2$ or ZrO$_2$ oxide scales containing high vapor pressure borosilicate phases. To satisfy the needs of long-term reusable hypersonic service environments, both matrix modification and deposition of coatings onto fibers and substrates are required, but challenges persist. Annular matrix cracking and thermal expansion mismatches between the C/C and coatings often lead to rupture during thermal cycling and oxidative erosion, both of which are issues for sustained high Mach number flight~\cite{zimmermann2008thermophysical}. \\

In an attempt to improve mechanical resilience and resistance to thermal shock, advancements in composite materials design have focused on multi-scale reinforcement strategies. Low dimensional micro/nanoscale reinforcements may include nanoparticles (0D), carbon nanotubes/fibers (CNTs/CNFs), whiskers (1D, e.g Si$_3$N$_4$, TaC, ZrC) or graphene (2D). Such additives serve to improve properties at the fiber-matrix interface via grain refinement, debonding, deformation, pullout, bridging, crack, and deflection mechanisms propagation~\cite{fu2022micro}. Higher dimensions, i.e., 2.5-D, 3-D, and 4-D, can be achieved by adding short fiber to the resin matrix or waving and winding fiber strands into hoop or braid-like structures~\cite{fu2022micro}. \\

The synergistic effects of complex multi-scale coating structural modification and fiber-matrix interface optimization need to be further developed to expand thermomechanical and erosion resistance properties for increasingly severe service environments. The incorporation of modern computational approaches, such as those developed in the national Materials Genome Initiative~\cite{de2019new}, and integrated design of the fiber-matrix interface with unique combinations of additives, may serve to improve materials performance. However, capabilities for high-fidelity modeling of complex architectures are still limited~\cite{marshall2014national, fahrenholtz2017ultra}. There remains a large gap in the comprehensive performance of oxidation/ablation-resistant modified C/Cs in standard databases. Available material properties are presently insufficient for design and scalability to relevant components such as aeroshells and leading edge~\cite{marshall2014national}.\\

\begin{figure*}[ht!]
    \centering
    \includegraphics[width=0.99\linewidth]{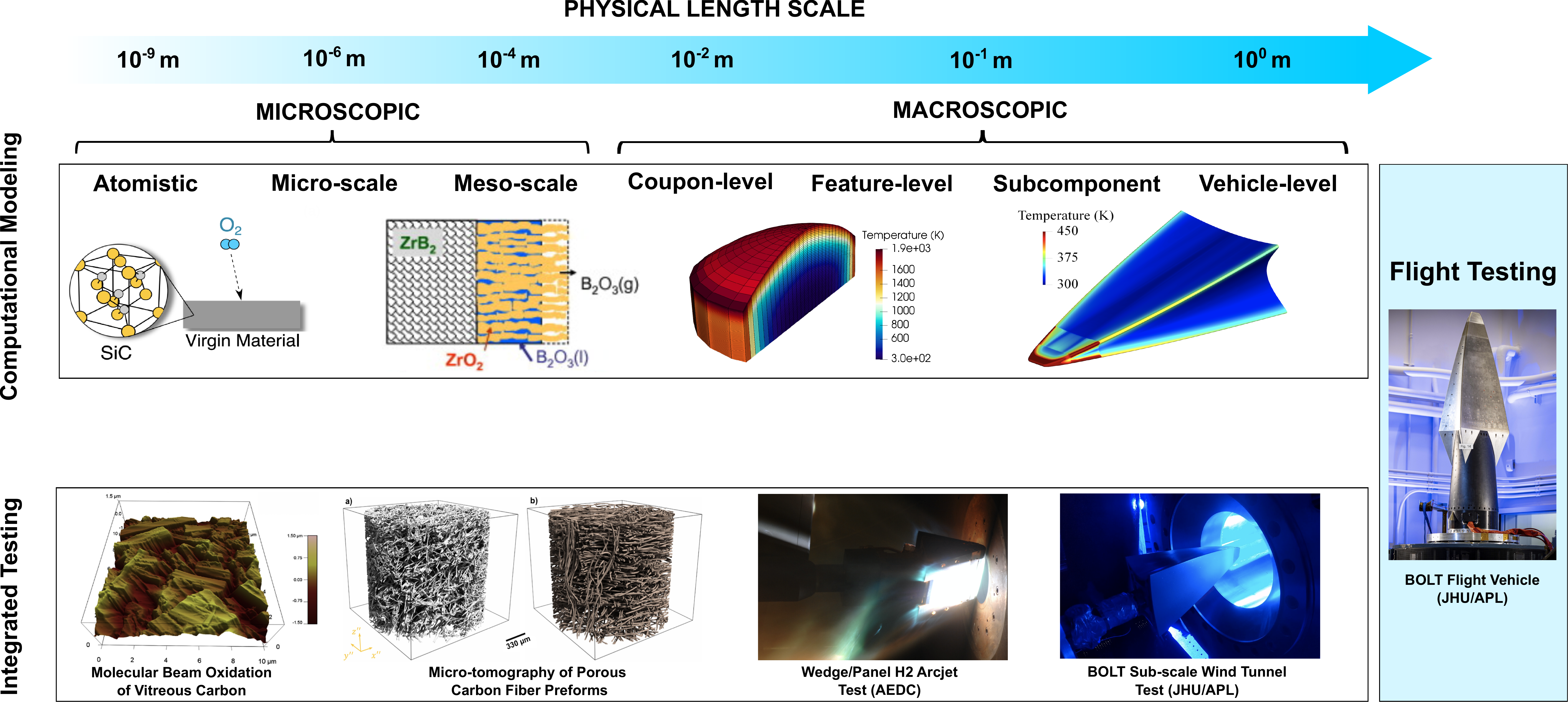}
    \caption{\small \textbf{Multi-scale modeling and testing framework for materials design and flight testing.} Length scales for both modeling and testing approaches span many orders of magnitude. Smaller-scale models inform and validate successively larger-scale tests. (Images adapted from~\cite{chen2020modeling,murray2020oxidation,parthasarathy2007model,panerai2017micro,wheaton2022} with permission.)}
    \label{fig:4}
\end{figure*}

\textbf{\small{Ultra-high Temperature and Refractory Ceramics for Hypersonics}.} Ultra-high temperature (UHT) and refractory ceramics are a developing class of materials for leading edges due to their stability at high temperatures~\cite {glass2011physical}. UHTCs, encompassing carbides, nitrides, and borides of early transition metals (Zr, Hf, Ti, Nb, Ta), possess high melting points (\textgreater 4000$^\circ$C), tunable densities (4.5-12.5 g/cm$^3$), high thermal conductivities \textgreater 140 W/mK), moderate coefficient of thermal expansion (6.3 to 8.6 x 10$^{-6}$/K), strong transition-metal-to-non-metal bonding (\textgreater 600 GPa mechanical stiffness), and high IR spectral emissivity for passive radiative cooling ~\cite{fahrenholtz2017ultra, wyatt2023ultra, ni2022advances}. The complexity of multiphase materials and their ability to survive extreme aerodynamic conditions is shown in \textcolor{blue}{Figure} \ref{fig:4}. Challenges concerning oxidation thermal shock can be mitigated by tailoring the architecture and leading to materials that rival metals in leading-edge applications. \\

Among the UHTCs, ZrB$_2$-SiC and HfB$_2$-SiC have received the most attention because they uniquely combine high thermal conductivity, specific strength ($\sigma$ $>$ 460 MPa at T=2500$^\circ$C, $\rho$=5.5 g/cm$^3$~\cite{zou2012high}), and superior oxidation resistance ${\sim}$1650$^\circ$C~\cite{zimmermann2008thermophysical, fahrenholtz2017ultra, mukherjee2018opportunities}. More recently, transition metal carbides have garnered interest as components for nozzle throats, divert/attitude control thrusters, and nozzle liners, where higher thermal and mechanical loads are encountered~\cite{glass2011physical}. Replacing HfB$_2$ and ZrB$_2$ with HfC or ZrC can increase service above 2000$^\circ$C. Still, the oxidation of refractory carbide and boride ceramics containing SiC remains a significant challenge for extended applications beyond ${\sim}$1600$^\circ$C. Above these temperatures, active oxidation generates gaseous oxidative products (i.e SiO$_{(g)}$ versus SiO$_{2(s)}$), which no longer provide protection against oxygen~\cite{glass2011physical, fahrenholtz2017ultra}.\\  

A certain level of porosity can improve resistance to thermal shock and thermal expansion mismatch. These pores help compensate for oxidation-induced volume expansion, leading to the formation of a fully dense and cohesive surface oxide scale but processing conditions and microstructure must be carefully controlled. The structure-processing property relationships for UHTCs are not well understood and additional information is needed to isolate fundamental factors that control the thermomechanical behavior of emerging compositions~\cite{fahrenholtz2017ultra}. For example, ZrB$_2$-MoSi$_2$ ceramics processed at elevated temperatures are noted to have improved oxidation resistance~\cite{silvestroni2017super}. \\

The best oxidation performance for monolithic ceramics is obtained using hot pressure assisted ceramics powder processing techniques, including hot pressing, hot isostatic pressing, and spark pressure sintering methods; these methods facilitate porosity reduction and sintering while limiting coarsening mechanisms~\cite{fahrenholtz2017ultra}.  Small UHTC grain sizes resist grain boundary fracturing during oxidation and restrict molecular oxygen transport, minimizing the disruptive effects of high-temperature martensitic transformations (phase changes) of HfO$_2$ and ZrO$_2$ (\textcolor{blue}{Figure} \ref{fig:4}). Other recent approaches for improved aerothermal resilience after evaporation of the protective B$_2$O$_3$ layer form from diboride oxidation include additions of W, Mo, and Nb~\cite{kazemzadeh2015effects} and graphene nanoplatelets reinforcement. These additions are indicated to suppress crack formation, bursting, and oxide layer by up to 60\%, while improving high heat-dissipative abilities, important to surface resiliency when exposed to plasma~\cite{nieto2014oxidation}. \\ 

\begin{figure*}[ht!]
    \centering
    \includegraphics[width=0.99\linewidth]{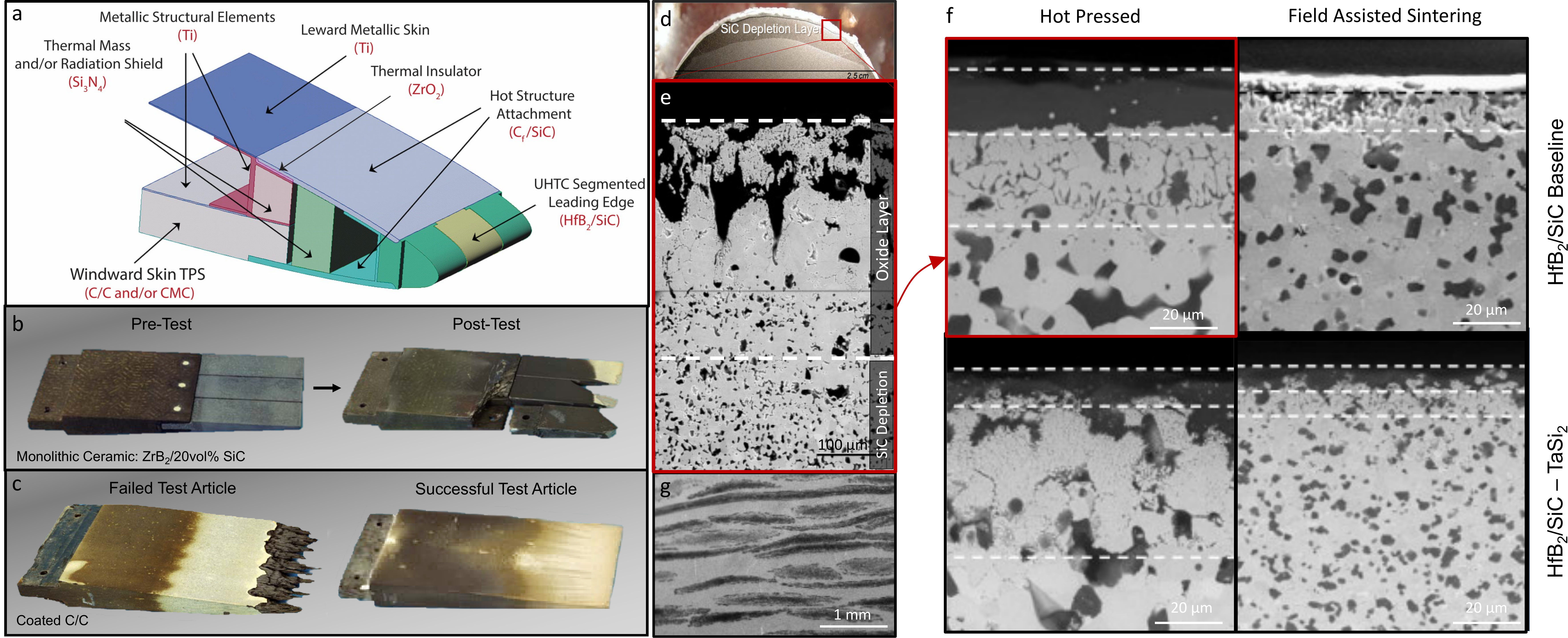}
    \caption{\small \textbf{Hypersonic wing leading edge designs and associated materials microstructures processed under using different conditions showing materials oxidation} {\bf (a)} Schematic drawings of wing leading-edge conceptual design using a monolithic ceramic segmented edge. {\bf (b)} Photographs of monolithic ZrB$_2$/20 vol\% SiC leading edges shown before and after arcjet testing in the H$_2$ arc-jet facility (\textcolor{blue}{Figure 4})   with failed ceramics due to oxidation and thermal shock. {\bf (c)} Failed and successfully coated-C/C X-43 leading edges following arc-jet for simulated flight conditions of Mach 10, 32 km altitude using 1475 W/cm$^2$, 130 seconds. {\bf (d)} HfB$_2$-SiC UHTC nose cone subjected to a total of 80 minutes of arc jet exposure at heat fluxes of 200 W/cm$^2$. The sample formed an oxide layer and a SiC depletion zone, which leaves behind a porous oxide surface {\bf (e)}. {\bf (f)} depicts SEM cross sections of HfB$_2$-SiC or HfB$_2$-SiC-TaSi$_2$ materials formed via hot pressing and/or field-assisted sintering and with or without TaSi$_2$ additives. The images indicate how grain structure, oxide layer formation, and SiC depletion are dramatically impacted by processing conditions and the inclusion of tertiary phases. {\bf (g)} SEM cross-section of a UHTCMC incorporating Cf and high aspect ratio SiC. (Images from NASA and adapted from~\cite{glass2008ceramic, glass2011physical, johnson2015ultra}.)}
    \label{fig:3}
\end{figure*}

The high densities of UHTC materials, low thermal shock resistance, and low fracture toughness impose additional physical limitations for bulk ceramics~\cite{glass2011physical}. Modern air-breathing hypersonic vehicles are extremely weight-sensitive. The high materials density (${\sim}$3-6 times the density compared to C/C) and poor thermal shock resistance (1/5 of ACC-6, and half that of RCC and CVI C/SiC 1100$^\circ$C~\cite{glass2011physical}) of monolithic ceramics become a limiting factor for structural components and dense segmented leading-edge inserts (\textcolor{blue}{Figure} \ref{fig:3}\textcolor{blue}{a, b}). As a result, the preferred instantiation of UHTCs is for emissive, anti-oxidative coatings on Cf composites or refractory alloys. UHTC coatings can be improved by adopting graded or layered compositions, enhancing bond strength by structural integration, enhancing toughness and crack bridging via nanoscale and micron-scale carbide fibers, and including emissivity enhancing dopants.   Compositional complexity can lead to further property improvements, where for example TaHf-C has the highest recorded melting temperature~\cite{cedillos2016investigating}. \\

Whether for monolithic ceramic bodies or barrier coatings, processing variables significantly affect material properties and impose difficulties in obtaining standardized performance data. Despite a significant body of research examining both practical densification mechanisms and compositional variation, limited information is reported on the kinetics of sintering and densification of bulk ceramics. Many studies have reported the thermomechanical properties of single ceramic compositions (strength, hardness, elastic constants, thermal conductivity, and fracture toughness), but limited information on the structure-processing-property relationship stemming from first principles are understood~\cite{fahrenholtz2017ultra, oses2018aflow}. To expand the utilization of refractory ceramics and UHTCs for hypersonic and extreme environments, the discovery/synthesis of candidate materials is needed, yet the probability of simple undiscovered compounds remains low~\cite{fahrenholtz2017ultra}. \\

Further advancement of materials design in these systems will likely parallel the development of structural metal alloys. Simultaneous improvement of oxidation resistance, creep at elevated temperatures, and transformation toughening will require additions of secondary, ternary additions, and high-entropy compositions that occupy uncharted areas of phase diagrams~\cite{marshall2014national, fahrenholtz2017ultra, ghosh2007integrated, oses2020high}. Moreover, recent advances in additive manufacturing has shown promise to produce UHTC components with tailored properties that might not be achieved using traditional spark plasma sintering or hot pressing techniques ~\cite{wyatt2023ultra, KEMP2021102049, PETERS2023103318, PETERS202311204}. Emerging work on RPME ceramic materials, such as (ZrHfTi)C solid-solutions is being conducted, where for example high Hf-content is beneficial for forming an amorphous oxycarbide layer enhancing initial oxidation resistance, while an equiatomic ratio of metallic atoms increased high-temperature phase stability~\cite{lun2021oxidation, cedillos2016investigating}. Multi-scale computational modeling of UHTCs can aid in the development of high-entropy materials by integrating ab-initio (fundamental chemistry and electronic properties), atomistic (thermomechanical properties), and continuum frameworks (mechanical properties, thermomechanical analysis of microstructure)~\cite{johnson2015ultra,toher2017combining, toher2014high, esters2021settling, esters2023qh}.  
Recent approaches have been established using a ``{\it disordered enthalpy-entropy descriptor}'' (DEED) to address functional synthesizability of high-entropy carbonitrides and borides~\cite{Nature_2024_DEED_Curtarolo_etal}.\\

\textbf{\small{Ceramics-Matrix Composites}.} The poor thermal shock resistance and high densities of bulk UHTCs and refractory ceramics can be overcome by the incorporation of ceramic fibers (${\sim}$35-60 vol\%) to create ceramics matrix composites (CMCs)~\cite{monteverde2017thermo, tang2017design, binner2019selection}. On the other hand, the oxidation resistance of C/C can be improved by replacing a carbonaceous with a ceramic matrix that forms a self-healing glassy oxide passivation layer. For the latter, SiC was determined to be a suitable substitute for the carbon matrix due to its high oxidation temperature, thermal shock stability, and creep resistance~\cite{krenkel2005carbon}. The most well-established carbon fiber-reinforced CMCs for hot structures are carbon fiber-reinforced silicon carbide composite (C/SiC), and carbon fiber-reinforced carbon-silicon carbide (C/C-SiC) composites~\cite{krenkel2005carbon}.  While the active oxidation of C/C starts at ${\sim}$500$^\circ$C in air and becomes more significant above 600$^\circ$C, C/SiC and C-C/SiC can be stable up to 1600$^\circ$C due to the formation of a thin self-passivating SiO$_2$ scale. Alternatively, environmentally stable oxide/oxide ceramic composites have been produced to combat oxidation experienced by non-oxide materials. These ``Ox/Ox'' CMCs incorporate polycrystalline alumina or aluminosilicate fibers (e.g Nextel 610 or Nextel 720) into alumina, aluminosilicate, or SiOC matrixes and were initially suggested for hypersonic thermal barrier materials. However, actual service applications are limited to temperatures ${\sim}$1000-1200$^\circ$C due to degradation in tensile strength, stiffness, and creep~\cite{volkmann2015assessment}.  \\

The fracture behavior of damage-tolerant CMCs is dominated by the stiff reinforcing Cf (or SiC$_f$), where fiber-matrix debonding is associated with frictional effects and crack deflection within porous or multilayer interfaces. Fiber/matrix bonding using a coating with adapted interphases (e.g. CVD pyrolytic carbon, silicon, $\beta$-SiC, BN, alumina) serves to (i) increase fiber-matrix bonding for enhanced mechanical properties (tensile strength ${\sim}$350 MPa for CVI C/SiC); (ii) protect carbonaceous fibers from oxidative degradation at ${\sim}$450$^\circ$C during crack formation when exposed to an oxidative atmosphere; and (3) mitigate the severity of CTE mismatch between C$_f$ and the SiC matrix in the absence of matrix damage~\cite{glass2011physical, krenkel2005carbon}. Due to the anisotropic nature of thermal expansion in composite materials and the CTE mismatch between the SiC matrix and the C$_f$, these materials may be more prone to cracking compared to SiC$_f$/SiC CMCs. In recent years, melt silicon carbide fiber-reinforced carbon-silicon carbide composites (SiC/SiC) can reach use temperatures up to 1600$^\circ$C. Yet, high temperature ``sweating'' of unreacted silicon leaves processing improvements to be desired. \\

Practically, similar manufacturing techniques for C/C fabrication can be used to infiltrate a carbon matrix with SiC or other refractory ceramic compositions (including UHTCs). CMC fabrication techniques include chemical vapor infiltration/deposition, PIP, reactive melt infiltration, slurry infiltration, in-situ reaction, hot pressing, and powder pre-infiltration. Several techniques can be combined to achieve multi-component compositions and gradient/sandwich structures. Still, the fabrication of complex ultra-high-temperature ceramic matrix composites (UHTCMCs) compositions remains of significant interest in replacing C/C or C/SiC CMC materials for improved thermomechanical capabilities. The advancement of UHTCMCs that incorporate HfC, ZrC, TaC, HfB$_2$, and ZrB$_2$ matrix compositions will serve to greatly benefit the development of propulsion platforms for hypersonic flights~\cite{tang2017design}. \\

Carbon fiber-reinforced UHTC-matrix composites as shown in \textcolor{blue}{Figure} \ref{fig:3}\textcolor{blue}{g}, especially those containing HfC and ZrC, can resist oxidation at temperatures above 2000$^\circ$C under hypersonic flight conditions. However, CTE mismatches between UHTC matrix phases are more prone to microcracking during processing and aerothermal heating that reduces strength~\cite{rueschhoff2020processing}.  The use of SiC fiber will improve the structural properties, and oxidation resistance (compared to C$_f$), and decrease the density of UHTCMCs. HfC composites reinforced with ${\sim}$15\% short linear chopped SiC fibers showed strengths up to~370 MPa at room temperature, decreasing to 290 MPa by 2200$^\circ$C in Ar~\cite{pienti2015microstructure}. These properties far exceed the strengths reported for C/C which has demonstrated strengths of ${\sim}$200 MPa at room temperature to 2200$^\circ$C. The addition of up to 40-60 vol\% SiC$_f$ is suggested to further improve mechanical properties~\cite{cheng2019flexural}. \\

\section*{\large{Advances in Computational Tools for Materials Development}} 
\textbf{\small{Advances in Theory and Computational Tools}.} Any experimental characterizations remain costly and/or inaccessible at the extreme conditions experienced by hypersonic vehicles~\cite{golla2020review, bale2013real, uyanna2020thermal}, making modeling/simulation efforts difficult to validate. Nonetheless, the lack of experimental data can be an opportunity for theory and computation to provide some insight. A history of simulation codes modeling thermal protection system materials is presented in~\cite{uyanna2020thermal} and dates to the 1960s and includes software like the CMA code by the Aerotherm Corporation~\cite{moyer1968analysis, chen1999ablation} and the FIAT code by the NASA Ames Research Center~\cite{chen1999ablation}, that incorporates internal energy balance, decomposition equations, general surface energy balance boundary conditions, and a thermochemical ablation model~\cite{uyanna2020thermal, riccio2017optimum}. Other simulations have been performed modeling ablation in carbon-phenolic (charring) composites~\cite{rivier2019ablative, wang2019assessment, mazzaracchio2018one}, electromagnetic shielding at microwave frequencies~\cite{albano2013electromagnetic}, water mass flow rate in transpiration (active) cooling systems~\cite{shen2016experimental}, flow fields and thermal behavior of solid samples, temperature-dependent fracture toughness of particulate-reinforced UHTCs~\cite{wang2017fracture}, and mechanical-thermoelectric performance for multi-functional thermal protection systems~\cite{gong2018novel}. High-temperature thermal and elastic properties of high-entropy borides have also been modeled using molecular dynamics~\cite{dai2021temperature}. All of these methods contribute to framing an understanding of the relevant vehicle systems level and materials design criteria. \\

The challenge of performing experimental characterizations at conditions relevant to hypersonic flight translates to a scarcity of empirical data. Computational data is similarly limited: the complex mass and heat transfer behavior at play spans several scales, and first-principles modeling is largely disjointed across these scales and far too expensive for high-throughput workflows. Any relevant data available exists in unstructured and programmatically inaccessible formats. This makes it difficult to leverage emerging artificial intelligence methods for accelerated screening. Solutions are on the horizon: synchrotron x-ray computed microtomography was able to fully resolve microcrack damage as cracks grew under load at temperatures up to 1750$^\circ$C~\cite{bale2013real} and high-throughput first-principles frameworks are becoming capable of accurately modeling finite-temperature properties~\cite{esters2021settling, esters2023qh, lederer2018search}. Disorder plays an ever-increasing role as the environment becomes more extreme~\cite{oses2020high, toher2019unavoidable}, further complicating characterization and modeling efforts. The prediction and optimization materials with useful properties will require an understanding of the interplay between ultra-high-temperature phenomena, which will be assisted by high-fidelity structured data, artificial intelligence, and solid thermodynamic-kinetic analysis. \\

To further the development of new and existing hypersonic materials, a computational approach alone is insufficient. Real-world data is needed to validate truly predictive material modeling and design. Furthermore, material phenomenology and behavior at extreme temperatures are difficult to predict by first principles alone, especially as material systems become more complex. \\


\textbf{\small{Role of Integrated Experiments and Flight Readiness Pathways}.} Within the material design framework, computational models and experiments serve vital and complementary roles. The intended application of the material being designed largely drives the environmental loads. For example, a TPS material designed for the exterior of a hypersonic vehicle will experience a vastly different heating and chemical environment than the inside of a scramjet combustor. These testing and modeling methodologies must take into account the coupled thermal, structural, and chemical nature of the material-environment interactions~\cite{chen2020modeling}. In addition, the intended reusability of the material dictates the relevant timescales that must be examined. For typical ballistic or boost-glide trajectories, flight times span tens of seconds to tens of minutes. A single-use application may be able to ignore slower phenomena that occur (e.g., creep), while a material designed for a reusable application must account for the overall life of the vehicle.\\

Computational models for materials are traditionally broken down by domain (material, fluid) and scale. At the larger scales, continuum codes rely on finite-element analysis (FEA) or similar numerical methods to solve the macroscopic governing equations over the physical domain. In the fluid domain, the focus is usually to predict the environment that the material experiences, such as the heat flux, pressure, and shear force. The field of computational fluid dynamics (CFD) has expanded to include thermal and chemical nonequilibrium, turbulence effects, surface reactions, and multiphase flow~\cite{candler2019, ching2020two}. On the material side, thermal and structural analyses have traditionally been considered separately. Material response tools evaluate the thermal and chemical response, including any ablation and pyrolysis within the material~\cite{amar2016overview}. Thermo-structural tools model the combined aero-mechanical and thermo-structural loads arising from thermal expansion and gradients within the materials.\\

In recent years, there has been a push towards higher-fidelity tools and more physics-based modeling, necessitating multi-scale modeling approaches to describe the material at smaller and smaller scales. An example of this is the focus on micro-scale modeling for porous carbonaceous ablative TPS materials by NASA~\cite{poovathingal2019nonequilibrium}. These span mesoscale models describing the distinct phases that are present in a material, down to atomistic models describing the fundamental material interactions. However, bridging the gap between the various scales using a truly physics-based approach is challenging given the relationship between materials processing, microstructure, physical properties, and thermal, structural, and chemical performance. These approaches can be generalized in an Integrated Computational Materials Engineering (ICME) framework. \\

Flight tests are prohibitively expensive, and this has historically been a major barrier to the development of hypersonic vehicles. Dedicated ground tests provide an alternative way to emulate flight conditions in a controlled environment. These include both aerothermal and structural tests. Although aerothermal ground tests seek to recreate flight conditions as accurately as possible, no facility can reproduce the exact flight conditions, and instead seek to match two or more parameters~\cite{kolesnikov2000concept}. Flight parameters of interest include (but are not limited to) Reynolds number, Mach number, heat flux, pressure, shear, temperature, chemical environment (e.g., dissociated air), thermal shock, and exposure time. The freestream enthalpies experienced during hypersonic flight are huge (in the MJ/kg), which presents another challenge for ground test facilities. Arc jets, which produce realistic flight enthalpies, shear, and pressure for up to several minutes, have been considered the gold standard of aerothermal testing for decades. However, these tests are costly and time-consuming to prepare and perform. Other facilities such as shock tubes/tunnels can match other flight parameters more accurately, but only for sub-second exposure times, which limits the utility of these facilities for thermal testing~\cite{holden2012review}. \\

Similarly, the aim of conventional thermo-mechanical ground testing is to validate some structural property, feature, or behavior given flight-realistic mechanical and thermo-structural loads. For external TPS materials, particularly C/C and ceramics matrix composites, properties such as the interlaminar and shear stress strengths are critical to the material performance and can vary greatly with respect to material processing. Thermo-mechanical testing can span simple property characterization, coupon-level, sub-scale, up to full vehicle-level tests~\cite{doug2017updated}. For materials designed for re-usable applications, the lifecycle of the material under the thermo-structural loads is also critical to evaluate, including any structural creep mechanisms. \\

For any flight vehicle, there is a large reliance on heritage materials (i.e., materials that have flown previously). For example, variants of low density alumina borosilicate insulation tiles (e.g AETB) that were originally roughly 50 years ago for the shuttle are still relied upon as a significant TPS modality for contemporary re-entry and hypersonic vehicles including Boeing X-37, Orion Multipurpose Crew Vehicle, SpaceX Starship, Stratolaunch Talon-A, and Sierra Space Dream Chaser ~\cite{stratolaunch} (\textcolor{blue}{Figure} \ref{fig:1}\textcolor{blue}{m}, \textcolor{blue}{Box 1}). Thus, there has historically been a large barrier to flight-test any candidate material system. In general, the technology readiness level (TRL) and the manufacturing readiness level (MRL) must be sufficiently high in order for a material to be considered flight-ready, depending on the risk tolerance of the flight program. Increasing both the TRL and MRL is a graduated process that requires extensive testing at multiple scales, ranging from coupon-level and sub-scale to full-vehicle tests. \\

\section*{\large{Outlook}} 

Each crucial sub-system within a hypersonic vehicle, including the primary structure, thermal protection, and propulsion system, must endure complex challenges encompassing aerothermal, oxidative, and extreme mechanical requirements. The advancement of these vehicles has been hindered by limitations in materials survivability, particularly in thermal protection and air-breathing propulsion systems. Overcoming these limitations is critical for fostering innovation in vehicle design. There is a pressing need to adopt expanded materials design frameworks, from atomistic to macroscopic, to propel the next generation of resilient materials for hypersonics. Some of the emerging efforts involve developing computationally informed materials design strategies to significantly enhance material properties, explicitly addressing pivotal challenges for refractory metals, composites, and refractive ceramic materials. We note some critical challenges and exciting opportunities in designing and implementing resilient materials for the next generation of hypersonic vehicles. \\

\textbf{\small{Metals for Hypersonics}} \\
Refractory metals, while possessing noteworthy attributes, exhibit limitations in oxidation resistance and strength at elevated temperatures, impacting their resilience in high heat flux environments. Future development of multi-principal element ``high-entropy'' alloys could provide significant benefits over legacy refractory alloys by decreasing density and improving oxidation kinetics. There remains a gap in our understanding of thermophysical properties in operation-relevant conditions. Leveraging rapidly evolving manufacturing methods alongside high-throughput characterization and machine learning algorithms holds promise for expeditiously uncovering novel compositions. Integrating these approaches could pave the way for novel advanced cooling technologies and resilient metallic structural components.

\textbf{\small{Composites for Hypersonics}} \\
Composites display promising strength-to-weight ratios and elevated temperature tolerance in inert atmospheres. Nevertheless, uncoated materials are susceptible to significant oxidation and erosion when subjected to extreme temperatures. Progress in enhancing performance will involve further improvements to applied high-temperature emissive protective coatings and the modification of carbonaceous matrices. Challenges arise from the disparity in material properties between coatings and matrices, as well as matrix-fiber interfaces. Improvements to thermomechanical models and simulations will be critical as present computational techniques struggle to scale properties from the microscale to the bulk.  The inherent anisotropic properties of existing fiber-reinforced materials underscore the importance of progress in multi-scale reinforcement strategies and higher-dimensional materials, which represent compelling avenues for future development.

\textbf{\small{Ceramics for Hypersonics}} \\
Refractory ceramics and UHTCs are characterized by exceedingly high melting points and thermal conductivities, but present challenges due to their low thermal shock resistance and density when used as monolithic components. Their optimal application lies in thermal barrier coatings and ceramics matrix composites. While transition metal carbides incorporating SiC have gained attention for components exposed to elevated thermal and mechanical loads, oxidation remains a formidable hurdle for extended applications beyond approximately 1600$^\circ$C. Continued research is imperative to exploring ceramics' structure-processing property relationships, transformation toughening, oxidation enhancement, and the intricacies of fiber/matrix bonding UHTCMCs.  Exploration of UTHCs in the domains of additive manufacturing, machine learning and modeling, and high-entropy UHTC compositions are poised to create complex new ceramics with tailored properties. Such methods will be critical in supporting hypersonic capabilities not previously achieved when using materials formed using conventional approaches.

\section*{Acknowldegements}
S.C. acknowledges support by DoD (N00014-21-1-2515) and NSF (NRT-HDR DGE-2022040). This material is based upon work supported by the Air Force Office of Scientific Research under award number FA9550-22-1-0221. I.M. and C.O. acknowledge support from the Air Force Office of Scientific Research, and discussions with Michael Brupbacher and Tom Magee.  T.M.P. acknowledges the support of a Department of Defense Vannevar Bush Fellowship Grant ONR N00014-18-1-3031.

\section*{Competing Interests}

The authors declare no competing interests.

\bibliography{mybibfile}

\end{document}